\documentstyle[prl,aps,epsf,multicol]{revtex}
\begin{document}
\draft
 
\title{Quantum Confinement Transition and Cuprate Criticality}
\author{T. Senthil and Matthew P. A. Fisher  
}
\address{ 
Institute for Theoretical Physics, University of California,
Santa Barbara, CA 93106--4030
}

\date{\today}
\maketitle

\def\i{\imath\,}
\def\ih{\frac{\imath}{2}\,}
\def\undertext#1{\vtop{\hbox{#1}\kern 1pt \hrule}}
\def\ra{\rightarrow}
\def\lfa{\leftarrow}
\def\ua{\uparrow}
\def\da{\downarrow}
\def\Ra{\Rightarrow}
\def\lra{\longrightarrow}
\def\ler{\leftrightarrow}
\def\lrb#1{\left(#1\right)}
\def\O#1{O\left(#1\right)}
\def\VEV#1{\left\langle\,#1\,\right\rangle}
\def\tr{\hbox{tr}\,}
\def\trb#1{\tr\lrb{#1}}
\def\dd#1{\frac{d}{d#1}}
\def\dbyd#1#2{\frac{d#1}{d#2}}
\def\pp#1{\frac{\partial}{\partial#1}}
\def\pbyp#1#2{\frac{\partial#1}{\partial#2}}
\def\ff#1{\frac{\delta}{\delta#1}}
\def\fbyf#1#2{\frac{\delta#1}{\delta#2}}
\def\pd#1{\partial_{#1}}
\def\br{\\ \nonumber & &}
\def\brr{\right. \\ \nonumber & &\left.}
\def\inv#1{\frac{1}{#1}}
\def\be{\begin{equation}}
\def\ee{\end{equation}}
\def\bea{\begin{eqnarray}}
\def\eea{\end{eqnarray}}
\def\ct#1{\cite{#1}}
\def\rf#1{(\ref{#1})}
\def\EXP#1{\exp\left(#1\right)}
\def\TEXP#1{\hat{T}\exp\left(#1\right)}
\def\INT#1#2{\int_{#1}^{#2}}
\def\MAT{{\it Mathematica }}
\def\LHS{left-hand side }
\def\RHS{right-hand side }
\def\COM#1#2{\left\lbrack #1\,,\,#2\right\rbrack}
\def\AC#1#2{\left\lbrace #1\,,\,#2\right\rbrace}

\begin{abstract}
Theoretical attempts to explain the origin
of high temperature superconductivity
are challenged by the complexity of the normal
state, which exhibits three regimes with increasing
hole doping:  a pseudo-gap regime when underdoped,
strange power laws near optimal doping and
more conventional metallic behavior when heavily overdoped.
We suggest that the origin of this behavior is linked to a
zero temperature quantum phase transition separating 
the overdoped Fermi liquid 
from a spin-charge separated underdoped phase.  Central to our
analysis is a new $Z_2$ 
gauge theory formulation, which supports topological vortex excitations - dubbed
visons.  The visons are gapped in the underdoped phase,
splitting the electron's charge and Fermi statistics into two separate excitations.
Superconductivity
occurs when the resulting charge $e$ boson condenses.  
The visons are condensed in the overdoped phase, thereby confining
the charge and statistics of the electron
leading to a Fermi liquid phase.
Right at the quantum confinement transition the visons are in a critical
state, leading to power law behavior
for both charge and spin.

\end{abstract}
\vspace{0.15cm}


\begin{multicols}{2}
\narrowtext

Despite the remarkable  
progress\cite{Timusk} in the experimental characterization of the 
cuprate high-$T_c$ materials,
a theoretical consensus on the important underlying physics
remains elusive.  Experiments have revealed a
rich phase diagram as the temperature and chemical doping
are varied, with low temperature spin and charge ordering in addition
to superconductivity.  The normal phase at elevated temperatures
is equally varied, exhibiting a pseudo-gap in the underdoped
regime and strange power laws at optimal doping.
In this paper, we propose a 
theoretical picture that provides a description of the basic aspects of all
parts of the cuprate phase diagram. 

\begin{figure}
\epsfxsize=3.5in
\centerline{\epsffile{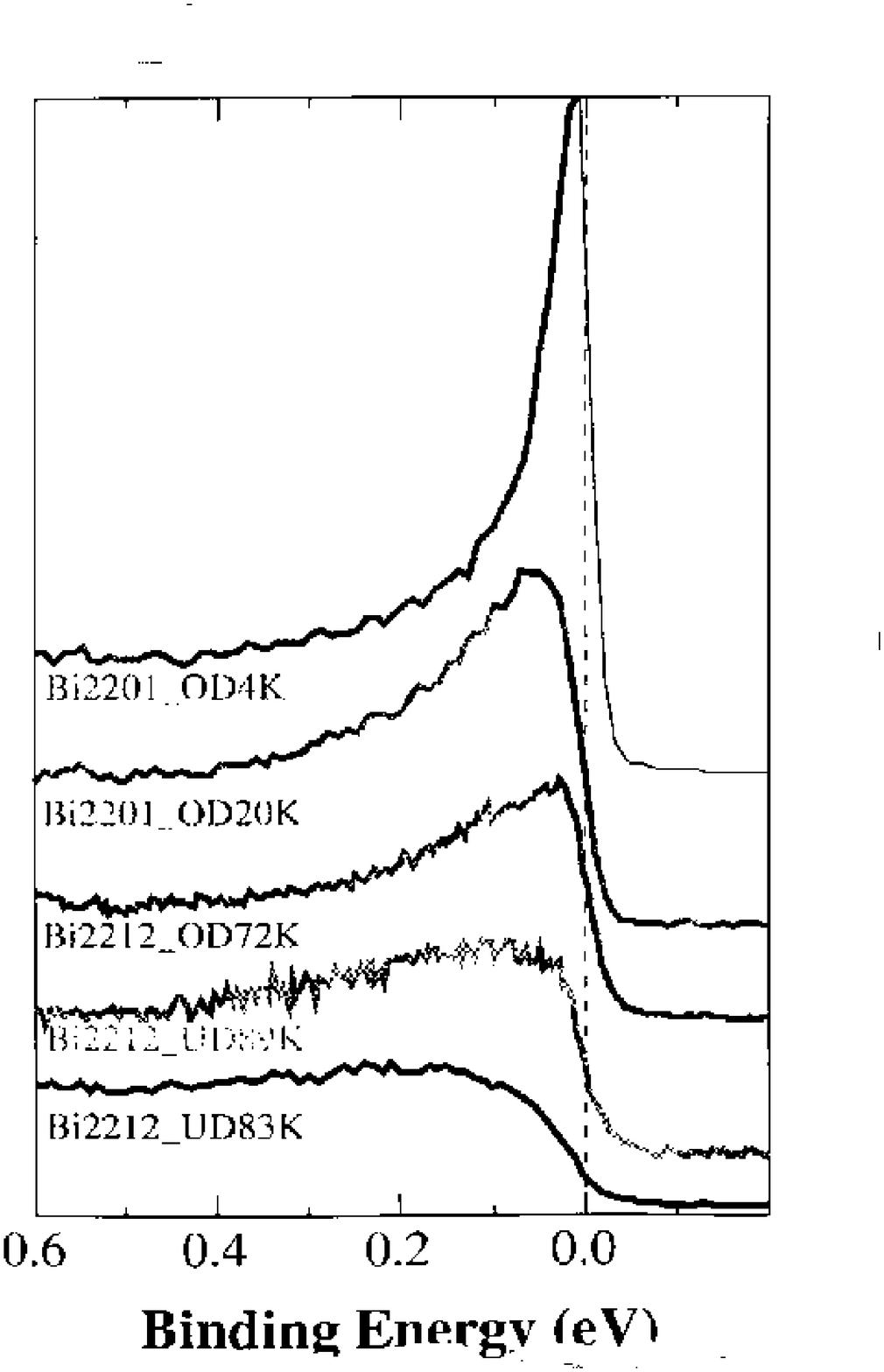}}
\vspace{0.15in}
\caption{Evolution of normal state ARPES lineshape with doping at momentum $(\pi, 0)$. The two lower curves are
for underdoped samples while the three upper curves are for overdoped samples. The data is from the 
group of J.C. Campuzano.}
\vspace{0.15in}
\label{camp}
\end{figure}

\begin{figure}
\epsfxsize=3.5in
\centerline{\epsffile{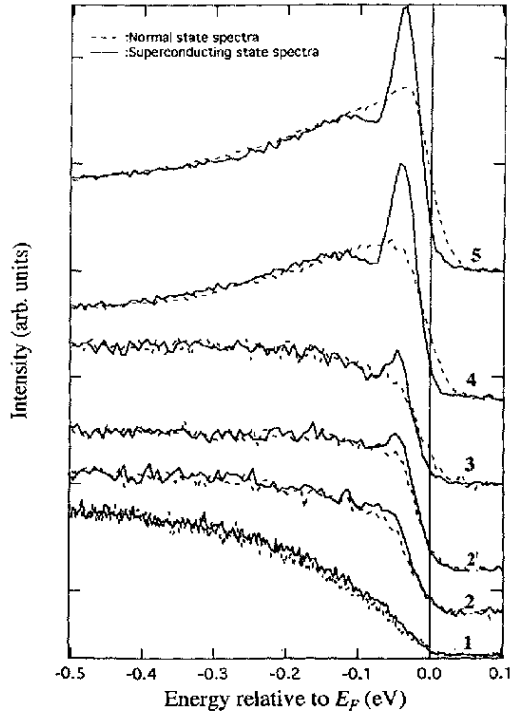}}
\vspace{0.15in}
\caption{ARPES spectra of BSCCO-2212 at momentum $(\pi, 0)$. The data is from the group of Z.X. Shen. 
The dashed lines are spectra in the
normal state, and the solid lines are in the same sample in the superconducting state. The highest curve 
corresponds to optimal doping with $T_c = 90 K$. The lower curves correspond to underdoped samples
with each successive curve corresponding to a lower value of $T_c$.}
\vspace{0.15in}
\label{shen}
\end{figure}

We begin with a brief discussion of experiment. In the last few years, angle-resolved photoemission
spectroscopy (ARPES) has emerged as an important experimental probe\cite{arpesrev} of the cuprates. 
The ARPES spectra provide a direct experimental measurement of the electron spectral function.
In {\em any} conventional phase (such as a Fermi liquid, band insulator, spin density wave or 
superconductor), a sharp quasiparticle peak is expected as a function of frequency ($\omega$)
at some momentum $\vec k$ in the Brillouin zone. The experimental results in the underdoped\cite{udarpes}
({\it and} undoped\cite{Shen}) cuprates in the non-superconducting state 
are in striking contrast to these expectations: the electron spectral function is highly smeared
with no trace of a sharp quasiparticle peak.  A sharp
peak does appear, however, upon cooling   
{\it into} the superconducting state\cite{Camp,Shen1}.  With increasing doping the normal state ARPES 
spectra sharpen somewhat, but even near  
optimal doping the observed peak is far too broad to be consistent with a 
conventional quasiparticle description. Some representative data may be found 
in Figs. \ref{camp} and \ref{shen}.

We take the absence of a quasiparticle peak in the ARPES data to be strong evidence that the electron
{\em decays} into other exotic excitations in the 
underdoped cuprates. Further evidence for this 
comes from transport measurements.  The $c$-axis d.c. resistivity 
shows ``insulating" behavior increasing
rapidly upon cooling, whereas the in-plane resistivity is 
typically ``metallic"
and much smaller in magnitude.
Moreover, in a.c. transport a Drude peak is observed
in the $ab$ plane, but not along the $c-$axis.
This strangely anisotropic behavior,
difficult to understand within 
a conventional framework, follows naturally
if the electron decays into exotic excitations
which reside primarily in the $ab$ plane. 
Transport along the $c$-axis requires hopping of {\it electrons} from layer to layer which is 
strongly suppressed
at low energies. 

If the electron indeed decays into other excitations,
what is their character?
There are two
distinct possibilities: (a) The electron may decay into two or more other exotic particles each of which carries some fraction of the quantum numbers of the electron (for instance, 
into separate spin and charge carrying excitations), or
(b) The exotic excitations may admit no ``particle'' description at all - this is known to happen generically  
at quantum critical points.  We hypothesize that (a) is realized in the underdoped cuprates. There are two reasons for doing so. First, the experiments strongly suggest that 
the electron decays throughout the underdoped region - 
fine-tuning to a critical point
as in possibility (b) appears unnecessary.  Second, 
as detailed below, the emergence of a 
sharp ARPES peak in the superconducting state points 
to the electron decaying into separate spin and charge carrying particle excitations.

In a recent paper\cite{z2g} we introduced a new theoretical formulation of 
strongly interacting electrons based on a $Z_2$
gauge theory, that enabled us to reliably demonstrate
the possibility of electron ``fractionalization" in two spatial dimensions.
The theory is closely linked to an earlier ``vortex field theory"
approach by Balents et. al.\cite{NL}, but is formulated in terms
of particle excitations - 
a charge $e$, spin $0$ boson (a chargon) and a charge $0$, spin $1/2$
fermion (a spinon), which are minimally coupled to a fluctuating $Z_2$ gauge field. Of particular importance to
issues of fractionalization are point-like vortex excitations
in this $Z_2$ gauge field, called ``visons".
Fractionalization is obtained whenever the visons are gapped.
When the 
visons {\it condense} the chargons and spinons
are {\it confined}, effectively ``glued" together to form an electron.
This results in a 
conventional phase where the excitations are electrons
(or electron composites such as a 
magnon or a Cooper pair).

The apparent decay of the electron in the underdoped cuprates,
strongly suggests that the visons
are gapped in this part of the phase diagram.  On the other hand, 
in the heavily overdoped region Fermi liquid behavior is expected,
implying a condensation of visons.  Together, this implies
that with 
increasing doping there must be a 
zero temperature phase transition where the visons first condense.
The existence of such a novel ``quantum confinement transition"
is essentially implied by the experimental data - the transition interpolates
between the deconfined underdoped region (with no quasiparticle peak)
and the heavily overdoped Fermi liquid regime.

\begin{figure}
\epsfxsize=3.5in
\centerline{\epsffile{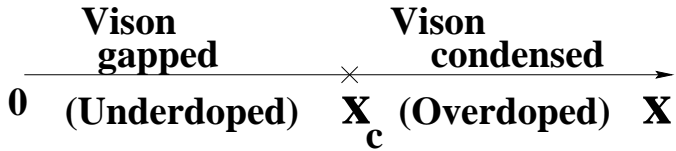}}
\vspace{0.15in}
\caption{Schematic zero temperature phase diagram as a function of doping $x$}
\vspace{0.15in}
\label{htct0}
\end{figure}

\begin{figure}
\epsfxsize=3.5in
\centerline{\epsffile{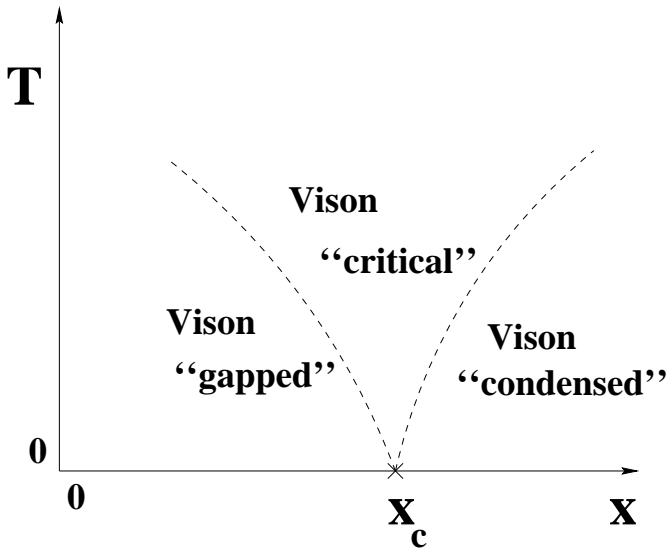}}
\vspace{0.15in}
\caption{Finite temperature crossovers in the vicinity of the 
quantum confinement transition at $x = x_c$ and $T=0$. The dashed lines denote crossovers, rather than finite temperature transitions.}
\vspace{0.15in}
\label{htct}
\end{figure}

A schematic zero temperature phase diagram paying attention only to the gross feature of 
whether the visons are gapped or condensed 
is shown in Fig. \ref{htct0}. Of particular interest is the quantum critical point associated with the 
confinement phase transition. It is clear that at finite temperature, the crossover from 
underdoped to overdoped regions will be determined by the properties of the quantum critical region associated with this quantum phase transition. 
In Fig. \ref{htct} we sketch the 
expected finite temperature crossovers in the vicinity of this phase transition.

The existence of a quantum confinement critical point controlling the crossover
from the underdoped to heavily overdoped regimes
is in qualitative agreement with a number of experiments.  It is well-known that this region is 
characterized experimentally by power-law temperature or frequency dependences of various physical 
quantities, as expected at a quantum critical point.
But more specifically, the sharpening of the ARPES spectra 
on moving from the underdoped to the overdoped region strongly suggests
that the  
critical point is associated with a {\it confinement} transition.

We now discuss the character of the two phases 
on either side of the confinement transition.
For $x>x_c$, and below the finite temperature crossover line, 
the system is presumably well decribed by Landau Fermi liquid theory.  In this
theory the low energy quasiparticle excitations near the Fermi surface are essentially electrons - they carry the
electron quantum numbers, spin 1/2 and charge $e$ - perhaps with
a renormalized effective mass.  As such, the electron spectral
function should exhibit sharp quasiparticle
peaks for all momenta lying on the Fermi surface,
which sharpen into delta functions at zero temperature.
Unfortunately, samples are difficult to grow in this heavily overdoped
regime, so that experimental data is rather limited.

But what is the character of the phase for $x<x_c$ where
the ``vison" excitations are gapped out?
This follows readily by inspecting the
effective action\cite{z2g} for the $Z_2$ gauge theory:
\bea
\label{IGA}
S & = & S_{c} + S_{s} + S_K + S_{B},  \\
S_{c} & = & - t_c \sum_{\langle ij \rangle} \sigma_{ij} ( b^*_ib_j + c.c.) ,\\
S_s &=& -\sum_{\langle ij\rangle} \sigma_{ij}
(t^s_{ij} \bar{f}_{i\alpha} f_{j\alpha} + t^{\Delta}_{ij} f_{i\ua}f_{j\da} + c.c ) - 
\sum_i \bar{f}_{i\alpha} f_{i\alpha} \\
S_K &=& -K \sum_{\Box} \prod_{\Box} \sigma_{ij}   .
\eea
Here, $b^\dagger_i$ creates a spinless, charge $e$ bosonic
excitation - the chargon - and $f^\dagger_i$ creates
the spinon, a fermion carrying spin $1/2$ but no charge.
When created together, these two excitations comprise the electron.
The field $\sigma_{ij}$ is a gauge field that lives
on the links of the space-time lattice, and takes on
two possible values:  $\sigma_{ij}=\pm 1$.  
The kinetic term for the gauge field, $S_K$,
is expressed in terms of plaquette products.  Here, $S_B$ is a Berry's phase\cite{z2g}
term which depends on the doping $x$.
The vison excitations are vortices in the $Z_2$ gauge field.
Specifically, consider the product of the gauge field $\sigma$ around an elementary plaquette, which can take on two values, plus or minus one.
When this product is negative, a vison excitation is present
on that plaquette.  Thus, when the visons are
gapped and absent in the ground state, all the plaquette
products equal plus one, and one can therefore
put $\sigma_{ij}=1$ on every link.  In this case the chargon
and spinon can propagate {\it independently}, and the electron
is {\it fractionalized}.  

\begin{figure}
\epsfxsize=3.5in
\centerline{\epsffile{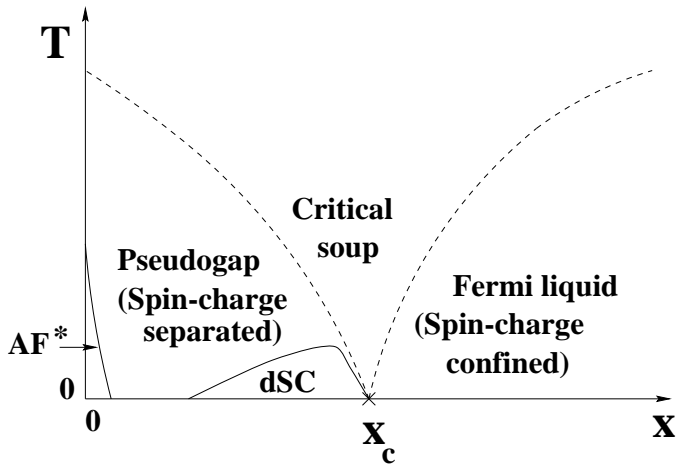}}
\vspace{0.15in}
\caption{Schematic finite temperature phase diagram as a function of doping $x$,
with the quantum confinement transition at $x=x_c$.
For $x$ just less than $x_c$ and at 
low temperature, the superconducting state arises as an {\em inevitable consequence}
of the fractionalization of the electron. At small $x$ and low temperature, the 
system orders antiferromagnetically. The resulting phase (denoted $AF^*$) is nevertheless spin-charge 
separated - see discussion in the text.}
\vspace{0.15in}
\label{htctsc}
\end{figure}

Once the electron is thus splintered, the character of the low temperature
phase will depend sensitively on the doping.  Based on knowledge of 
bosonic systems, one expects that the chargons will
condense into a superconducting phase upon cooling,
with $\langle b^\dagger \rangle \ne 0$.  But this condensation
can be easily impeded by commensurability effects 
from the underlying Copper-Oxygen lattice acting in concert
with the long-ranged Coulomb interaction.  Specifically, in the undoped limit with $x=0$ there is one charge $e$ chargon per unit cell,
and the chargons are expected to lock into a Mott insulating phase,
rather than condensing.  For very small $x$ with the doped
holes well separated, the chargon motion will still
be greatly impeded by the near commensurability,
and the long ranged Coulomb interactions should drive
charge ordering into an insulating state.  Thus, one only
expects the chargons to condense into a superconducting phase
for $x$ just less than $x_c$, as depicted schematically
in Fig. \ref{htctsc}, and consistent with experiment.

What is the nature of the chargon condensed superconducting phase?
In a conventional BCS description of superconductivity,
two electrons near the Fermi surface pair, and the resulting
charge $2e$ Cooper pairs condense.  
Within this charge $2e$ boson condensed superconductor,
the flux quantum is halved, given by
$\phi_0 = (1/2) (hc/e) = hc/2e$.  This is the value of the observed
flux quantum in the Cuprate superconductors, suggesting that
the superconducting phase itself is of the BCS variety.  But the
chargons carry the electron charge $e$, so one might have thought
that the BCS superconductor would be equivalent
to a chargon-pair condensate - with $\langle b^2 \rangle \ne 0$
yet $\langle b \rangle =0$ - rather than a single
chargon condensate, with $\langle b \rangle \ne 0$.  But, quite remarkably,
this is not the case.  As detailed in Ref. \cite{z2g}, it is the condensation
of single charge $e$ chargons that corresponds to the conventional
BCS superconductor, whereas the chargon pair condensate describes
an exotic non-BCS superconducting phase. 
 
This remarkable fact indicates a new route to superconductivity,
very different from a Cooper pairing of electrons.  Instead,
via a fractionalization process the electron charge
is liberated from it's Fermi statistics - resulting
in bosonic charge $e$ particles.  A direct condensation of
these chargons gives the conventional BCS superconducting phase.
Since fractionalization is tantamount to a gapping of the vison
excitations, this occurs below the crossover line depicted in Fig. \ref{htctsc}.
Thus, below this crossover line one has ``preformed"
superconductivity, with liberated chargons poised
to condense.  The electron spin is carried by fermionic spinons
in this regime, which are presumed to be gapped throughout the
Brillouin zone, except for four gapless nodal points.  This leads
naturally to a gapping of spin excitations upon fractionalization.
Thus, the non-superconducting vison ``gapped" regime
can account for the observed ``pseudo-gap" phase
in the underdoped cuprates.

Finally, we discuss the regime intervening between
the pseudo-gap and Fermi-liquid phases, centered
around $x=x_c$.  In this regime the visons are
neither gapped nor condensed, but in a
critical state.  The chargons and spinons
which are separated in the vison gapped regime, and confined
into the electron when the visons have condensed, are
in a state of limbo near $x = x_c$.  They cannot move
as independent free excitaitons since they are both strongly coupled to the
critical fluctutations of the visons, but they also cannot
move together as a confined electron.  The precise behavior
in this critical regime will be controlled by the nature
of the zero temperature quantum phase transition,
at $x=x_c$ in the Fig. \ref{htct0}. 

\begin{figure}
\epsfxsize=3.5in
\centerline{\epsffile{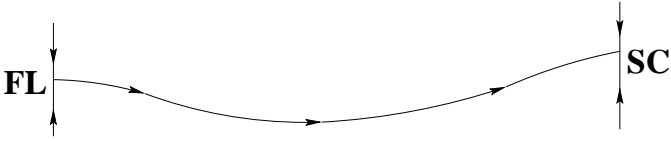}}
\vspace{0.15in}
\caption{A two dimensional section of the renormalization group flow diagram showing the instability of a Fermi liquid 
in the presence of arbitrary weak attractive interactions. The resulting Cooper pairs condense,    
leading to superconductivity.
}
\vspace{0.15in}
\label{bcsrg}
\end{figure}

\begin{figure}
\epsfxsize=3.5in
\centerline{\epsffile{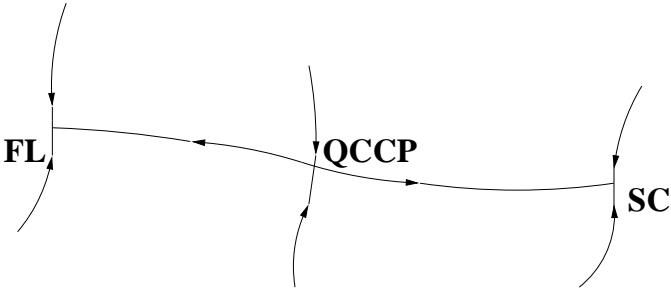}}
\vspace{0.15in}
\caption{A two dimensional section of the renormalization group flow diagram 
illustrating the different route to 
superconductivity envisaged in this paper. Central to the proposal is the existence of 
an unstable fixed point(QCCP) controlling a quantum phase transition 
at the point of instability of a 
Fermi liquid toward fractionalization. The resulting spinless charge $e$ bosons condense,
leading to superconductivity.}
\vspace{0.15in}
\label{qccprg}
\end{figure}

To our knowledge, the possibility and implications of a direct quantum phase
transition between a d-wave superconductor and a Fermi liquid
phase has not been discussed previously.  Within conventional BCS theory there
is {\it no} quantum phase transition separating the Fermi liquid
and superconducting phases.  Rather, the Fermi liquid
phase in the presence of arbitrarily weak phonon mediated
attraction between the electrons is {\it unstable} to the
formation of Cooper pairs which then condense leading to 
superconductivity.  Within a modern renormalization group
framework, one would say that the fixed point describing the
Fermi liquid phase is unstable and crosses over to the
superconducting fixed point, as depicted schematically in Fig. \ref{bcsrg}.
We are suggesting an alternate possibility interconnecting
these two phases.  As depicted in Fig. \ref{qccprg}, we imagine the existence
of an unstable fixed point, denoted QCCP(quantum confinement critical point), 
which controls
the nature of a strong coupling zero temperature phase transition
between the Fermi liquid and superconducting phases.   
The existence of such
a fixed point is strongly implied by our $Z_2$ gauge theory formulation.
To see this, imagine initially decoupling the chargons and spinons
in Eqn. \ref{IGA}, by setting $t_c=t_s=t_{\Delta} =0$ and putting $S_B=0$.  The remaining theory
describes a pure $Z_2$ gauge theory, which has two phases\cite{Wegner} - 
a phase with gapped visons for $K>K_c$, and a vison
condensed phase when $K<K_c$.
Now recouple the chargon and spinon fields.  When the visons
are gapped, the chargons and spinons can propagate
independently, forming a Bose and Fermi fluid,
respectively.   Presuming one is not too close
to a strongly commensurate filling, the fluid of bosonic
chargons should condense at low temperatures
giving superconductivity.  On the other hand, when the {\it visons}
condense, they {\it confine} the spinons and chargons,
giving fermionic charge $e$ carriers - the electron. 
Forming a fermionic fluid, these electrons of course cannot condense.
Rather, one expects that away from commensurate fillings they will form a conventional metallic Fermi liquid.  Finally, right at $K=K_c$
the visons will be in a critical state - described\cite{Wegner}
by the classical 3d Ising model when the spinon
and chargon coupling is ignored.  Here, one expects the spinons and chargons to be strongly scattering off these critical fluctuations, forming
a strongly interacting ``soup".

We now turn briefly to a few experimental implications of
the above scenario, focussing initially on the vison gapped
regime for $x<x_c$.  Here, the electron is fractionalized - 
an electron added to the system will decay into
a spinon and chargon.  This has direct implications for
electron photoemission experiments.  Since the electron decays
one does not expect a sharp spectral feature in photoemission.
More formally, in this regime the electron propogator, $G(r,\tau)$, can be roughly expressed
as a product of the chargon and spinon propogators, $G_c$ and $G_s$:
\begin{equation}
\label{elgf}
G(r, \tau) \approx G_c(r, \tau) G_s(r, \tau)  .
\end{equation}
The spectral functions for the spinons and chargons ($A(k,\omega) = -\frac{1}{\pi}Im G(k,\omega)$)
will have sharp spectral features since these particles can propagate
coherently when the visons are gapped, but the {\it electron}
spectral function is a convolution of these two and will hence not
exhibit any sharp spectral features.  This is exactly as seen in the normal state ARPES spectra in the underdoped samples. 
Now consider cooling the system into the superconducting state. As explained above, 
this requires condensation of the chargons so that
\be
G_c(r, \tau) \approx |< b >|^2  .
\ee
Then, from Eqn. \ref{elgf}, the electron Green's function just reduces to
\be
\label{elgfsc}
G(r, \tau) \approx |<b>|^2 G_f(r, \tau)  ,
\ee
and is simply 
proportional to the spinon Green's function inside the superconductor. 
Since the spinons propagate coherently, a sharp quasiparticle peak is expected - exactly as seen in the experiments. 
Moreover, 
since the {\it amplitude} of the peak is 
proportional to $| < b > |^2$,  it should become smaller as the 
superconductivity weakens, for instance, by reducing the doping. 
This is also borne out by the 
photoemission data\cite{Shen1} - see Fig. \ref{shen}. Thus, the qualitative trends in the underdoped
photoemission experiments can be 
well explained by assuming the electron decays into a chargon and a spinon.

For $x > x_c$, the low temperature properties of the system should be those of a Fermi 
liquid. This is commonly believed to be true.
It would, however, be useful to have 
more detailed experimental support. 

Now consider the ``quantum critical'' regime with $x \approx x_c$. As is usual near critical points,
power law temperature dependences are expected for various physical quantities. It is well-known that 
this is seen in a variety of experiments near optimal doping. In particular, the resistivity in the $ab$ 
plane exhibits a striking linear temperature dependance. In our scenario, the scattering of the chargons off
the critical visons is expected to give a power law resistivity $\rho(T) \sim T^{\alpha}$ with an 
exponent $\alpha$ that is at present unknown. Calculation of this and other universal properties of this 
quantum confinement transition is an important challenge that we leave for future work.

Thus far we have primarily focussed on the doping regimes near the superconducting phase.
We now turn to the {\it highly} underdoped and undoped materials. As discussed previously, 
upon approaching half-filling the condensation of the chargons is expected to be inhibited by 
commensurability effects together with the long-range Coulomb interactions. Instead, the chargons 
will localize. Away from half-filing, the charge localization will break the translational symmetry of 
the lattice. This is qualitatively consistent with the several experiments that observe stripe formation
in this region at low temperatures. It is important to stress, however, that in our scenario charge localization 
and translational symmetry breaking {\em coexist} with 
electron fractionalization. 

What is the fate of the gapped visons in the undoped material? It is very well 
established that the undoped insulator has antiferromagnetic long-ranged order. 
But magnetic order, just like charge order,
is conceptually independent of whether or not 
the electron is fractionalized, in other words, whether the vison is gapped or not. One can therefore 
contemplate two possibilities - (a) the visons are gapped in the undoped antiferromagnetic 
insulator, denoted $AF^*$, or 
(b) the visons are condensed leading to a conventional antiferromagnetic insulator,
denoted $AF$, with the electron in the spectrum. 
Note that the excellent description of the {\it low energy} spin physics by the quantum Heisenberg
spin model is not sufficient to dispose of this question. 
In fact, the two alternatives are distinguished by the nature of
the {\it gapped} excitations.  Experimental evidence for possibility
(a) follows from recent photoemission experiments\cite{Shen} on undoped
cuprates,
which do not exhibit a 
sharp quasiparticle peak 
at {\em any} momentum in the Brillouin zone.   
Following Balents et. al.\cite{NL}, we
thus suggest that that the electron
decays even in the undoped material, and that the visons are gapped. 
Further qualitative support 
is provided by mid-infrared optical absorption\cite{iroa} and Raman\cite{raman} 
measurements in the undoped material
which exhibit broad spectral features out to rather high energies,
not expected for the simple Heisenberg model.

Since the original discovery of high temperature superconductivity, literally thousands of theories
have been put forward to explain the phenomena.  The scenario we describe above has some overlap
with many earlier approaches, but is perhaps closest in spirit to the original Resonating
Valence Bond (RVB) theory of Anderson\cite{PWA}.  Here, we briefly mention the key similarities and differences
with the RVB theory.  In the original RVB theory,
spin-charge separation is intimately connected with
the presence of a ``spin-liquid" Mott insulating state, which was argued
to support neutral spin one-half spinon excitations.
It was soon established, however, that the undoped
parent compounds are not spin-liquids but rather antiferromagnetically
ordered.  It then appeared that the RVB state, if present at all, required the presence of doped holes.
In sharp contrast, within our $Z_2$ gauge theory approach spin-charge separation - or more generally
electron fractionalization - is {\it not} directly linked to magnetic ordering.
Rather, electron fractionalization occurs whenever the visons are gapped.
This is possible even in the presence of long-range magnetic order, in which case
the gapless magnons co-exist with gapped spinon excitations.  We believe that this is a likely
situation in the undoped cuprates.  

The original motivation for the RVB approach was based on an analogy
with the physics of spinons in one-dimension.  But our approach
demonstrates that spin-charge separation in two-dimensions requires the existence of a deconfined
phase of the underlying $Z_2$ gauge theory.  In one-dimension
this gauge theory always confines, and spin-charge separation occurs via a different solitonic mechanism.  
Apparently, this solitonic mechanism
of spin-charge separation encapsulated within the RVB approach\cite{KRS},
is not generally operative in higher dimensions.
In more formal terms, one can attempt\cite{oldgauge} to implement
RVB theory directly in two-dimensions with a $U(1)$ or $SU(2)$ gauge theory.  
But despite the apparent similarity with our $Z_2$ gauge theory, these continuous gauge theories
do not have a deconfined phase, and are thus apparently incapable of describing
spin-charge separation.

Within the RVB framework, it is common to describe the valence bonds
as being ``Cooper pairs" pre-formed in the insulator\cite{PWA},
which become mobile upon doping and condense into
the superconducting phase.  In contrast, in our $Z_2$ gauge theory the picture 
underlying the superconductivity
is the liberation of the electron's charge from
it's Fermi statistics, to form bosonic charge $e$ particles - the chargons.
Upon doping the chargons become mobile and can condense giving rise to superconductivity.

An entirely new aspect of our approach is the suggestion of a quantum confinement transition
separating the spin-charge separated pseudo-gap regime from the heavily overdoped
Fermi liquid phase.  At this transition the visons are neither gapped nor condensed, 
but rather in a gapless critical state.  Similarly, the spinons and chargons can neither propagate
coherently as independent excitations nor as a confined electron.  
We believe that this quantum confinement
transition might well account for much of the novel behavior observed near optimal
doping in the cuprates.  Developing a theoretical approach to access the properties
of such confinement transitions remains as an important yet challenging task.

We are grateful to 
L. Balents, G. Baskaran, C. Lannert, P.A. Lee, T.V. Ramakrishnan, G. Sawatzky,
R.R.P. Singh,
and Doug Scalapino 
for illuminating conversations.
We would especially like to thank Chetan Nayak for
emphasizing to us the importance of the crossover
to the heavily overdoped portion of the cuprate
phase diagram, and J.C. Campuzano and Z.X. Shen for 
permission to reproduce their experimental data. 
This research was generously supported by the NSF 
under Grants DMR-97-04005,
DMR95-28578
and PHY94-07194.

\end{multicols} 
\end{document}